\documentclass{article}
\usepackage[preprint]{neurips_2026} 

\usepackage[utf8]{inputenc} 
\usepackage[T1]{fontenc}    
\DeclareUnicodeCharacter{2013}{--}
\DeclareUnicodeCharacter{201C}{``}
\DeclareUnicodeCharacter{201D}{''}
\usepackage{textcomp}
\usepackage{hyperref}        
\usepackage{url}            
\usepackage{booktabs}       
\usepackage{bbm}           
\usepackage{xcolor}
\usepackage{amsfonts}       
\usepackage{nicefrac}       
\usepackage{microtype}      
\usepackage{multirow}
\usepackage{enumitem}
\usepackage{amsmath}
\usepackage{graphicx}
\usepackage{array}
\usepackage{wrapfig}
\usepackage{amsthm}

\theoremstyle{plain}
\newtheorem{theorem}{Theorem}[section]

\theoremstyle{definition}
\newtheorem{definition}[theorem]{Definition}

\theoremstyle{remark}

\theoremstyle{definition}  
\newtheorem{problem}[theorem]{Problem}

\newcolumntype{C}[1]{>{\centering\arraybackslash}p{#1}}
\newcolumntype{L}[1]{>{\raggedright\arraybackslash}p{#1}}
\graphicspath{{figures/}}
\usepackage{makecell}

\providecolor{orange}{RGB}{230,126,34}
\definecolor{answerorange}{RGB}{230,126,34}

\newcommand{\missingasset}[1]{%
  \fbox{\begin{minipage}[c][0.18\textheight][c]{0.90\linewidth}
  \centering Missing asset: \texttt{\detokenize{#1}}
  \end{minipage}}%
}
\newcommand{\safeincludegraphics}[2][]{%
  \IfFileExists{figures/#2}{\includegraphics[#1]{#2}}{%
    \IfFileExists{#2}{\includegraphics[#1]{#2}}{\missingasset{#2}}%
  }%
}

\hypersetup{
  colorlinks=false,
  pdfborder={0 0 1}
}

\title{A Nationwide Benchmark for Wildfire Initial Attack Failure Prediction with Public Environmental Data}

\author{
Runyang Xu\textsuperscript{1} \\
\textsuperscript{}Florida State University \\
\texttt{rx23@fsu.edu}
\And
Xueqi Cheng\textsuperscript{1} \\
\textsuperscript{}Florida State University \\
\texttt{xc25@fsu.edu}
\And
Yushun Dong\textsuperscript{}\thanks{Corresponding author} \\
\textsuperscript{}Florida State University \\
\texttt{yushun.dong@fsu.edu}
}

\newcommand{\benchmarkname}{\textsc{WildfireIA}}

\begin{document}
\maketitle

\begin{abstract}
Initial attack (IA) is the first wildfire suppression phase, when agencies must quickly decide which fires may escape early control. 
Existing IA failure prediction studies often use non-public response records or regional settings, so it remains unclear how well public data available at fire discovery time can support IA failure prediction at national scale. 
We present \benchmarkname{}, the first U.S. national-scale benchmark for IA failure prediction from environmental and contextual data available at fire discovery time. 
\benchmarkname{} aligns 38,128 naturally caused FPA-FOD wildfire events with FIRMS/VIIRS thermal detections, gridMET weather and fire-danger variables, LANDFIRE vegetation, fuel, and topography, OpenStreetMap access features, and WorldPop population density. 
To prevent data leakage, the benchmark fixes the event unit, size-based label rule, chronological split, metrics, and forbidden-feature list, and excludes final fire size, containment timestamps, and post-discovery satellite detections from model inputs. 
We evaluate 16 representative models across tabular, temporal, spatial, and spatiotemporal families under the same protocol. 
Results show that public discovery-time data provides useful but incomplete signal for IA failure prediction: XGBoost achieves the best AUPRC of 53.3\%; FIRMS/VIIRS is the least redundant source; and fuel is the strongest static predictor when dynamic observations are unavailable. 
We release preprocessing outputs and model-ready caches to support reproducible research on early wildfire risk assessment: \url{https://github.com/LabRAI/WildfireIA#}.
\end{abstract}

\section{Introduction}\label{sec:intro}

The initial attack (IA) phase is the first suppression phase after a wildfire is discovered, when first-arriving firefighting resources attempt to halt or slow fire spread~\cite{lee2013deploying,korkola2024ia}. 
At this stage, fire management agencies face an event-level triage problem: they must quickly predict whether a reported wildfire can be contained by the initial response or is likely to escape early control~\cite{korkola2024ia,cardil2025suppression,xu2023initialattack,katuwal2018predictattack}. If this risk is underestimated, a wildfire that still appears manageable may be treated just as a routine event, causing agencies to miss the narrow window for early escalation and forcing a shift from rapid containment to a larger, longer, and more costly suppression effort with greater operational complexity and life-safety risk~\cite{katuwal2018predictattack,rashidi2018attacker,cardil2025suppression,arienti2006empirical}. These stakes are becoming increasingly severe in U.S. recently as fire-weather seasons lengthen, large-fire activity increases, fuels become drier, and ignition patterns expand the conditions under which damaging wildfires can occur~\cite{westerling2006warming,jolly2015climate,dennison2014large,abatzoglou2016climate,balch2017human}.

Previous IA failure prediction studies show that early suppression outcomes are associated with fire weather, fuels, terrain, access, location, and response-related factors~\cite{arienti2006empirical,lee2013deploying,korkola2024ia,cardil2025suppression}. 
However, these studies leave two immediate barriers to building reliable early-warning models. 
First, many rely on operational variables such as dispatch time, crew size, resource availability, aircraft use, and deployment logs, which are important for explaining suppression outcomes but are often unavailable in public U.S. national databases or unknown at fire discovery time~\cite{lee2013deploying,katuwal2018predictattack,rashidi2018attacker,korkola2024ia}. 
This makes their results difficult to reproduce and can mix the intrinsic early risk of a fire with information about the later suppression response. 
Second, existing evaluations are often regional or agency-specific, so their labels, response standards, source availability, and evaluation protocols are not directly comparable across studies~\cite{korkola2024ia,butler2025victoria,xu2023initialattack}. 
As a result, agencies and researchers still lack clear evidence on how well public discovery-time data can predict IA failure, which sources add non-redundant signal, and which models are reliable under the same evaluation setting, limiting evidence-based early warning and resource prioritization. 
This uncertainty has practical consequences as wildfire activity continues to strain communities and response systems, with U.S. fires in 2026 already exceeding the 10-year average for the same year-to-date (YTD) period in both fire counts and burned area, and recent evidence showing that wildfire smoke is imposing growing public-health burdens~\cite{nifc2026nationalfire,deng2026fireozone}. 
Therefore, a U.S. national public benchmark is urgently needed to turn IA failure prediction into a reproducible, leakage-controlled task where data sources and model families are compared under the same rules.

To tackle the research gaps, in this paper we present \benchmarkname{}, the first U.S. national-scale benchmark for wildfire IA failure prediction from public environmental data available at fire discovery time. To address the dependence on non-public operational records, \benchmarkname{} builds on 38,128 naturally caused FPA-FOD wildfire events as a national event backbone and aligns them with public discovery-time or static signals from FIRMS/VIIRS, gridMET, LANDFIRE, OpenStreetMap, and WorldPop~\cite{short2017fpafod,nasa_firms,abatzoglou2013gridmet,landfire,osm,worldpop}. To address the lack of comparable evaluation protocols, \benchmarkname{} fixes the event unit, size-based label rule, chronological split, forbidden-feature list, metrics, and model-ready representations, while excluding outcome-derived information such as final fire size, containment timestamps, and post-discovery satellite detections from all model inputs~\cite{short2017fpafod,xu2023initialattack,korkola2024ia,bhardwaj2025containment,mtbs}. 
We focus on naturally caused wildfires because their early spread is more directly governed by ignition environment, fire weather, fuel, and terrain, whereas human-caused fire outcomes are more strongly shaped by human activity distributions and reporting patterns~\cite{balch2017human,finney1998farsite,finney2006flammap,jolly2015climate}. 
Using this benchmark, we evaluate 16 representative models that cover both standard classical baselines and recent state-of-the-art architectures across tabular, temporal, spatial, and spatiotemporal prediction families~\cite{chen2016xgboost,cho2014learning,shi2015convlstm,tang2023swinlstm,garnot2021utae}.
The evaluation shows three key insights: (i) public discovery-time data contains meaningful but incomplete signal for IA failure prediction, with structured event-level models providing strong baselines; (ii) discovery-day FIRMS/VIIRS evidence is the least redundant input source, while fuel provides the strongest static fallback signal when dynamic observations are unavailable; and (iii) containment duration is only weakly explained by discovery-time inputs, indicating that post-discovery suppression actions and fire evolution remain dominant factors. Overall, our contributions are summarized as follows:

\begin{itemize}[leftmargin=*]
    \item \textbf{The first U.S. national benchmark for wildfire IA failure prediction.}
    We introduce \benchmarkname{}, which standardizes IA failure prediction as an event-level task at fire discovery time, with fixed event units, labels, chronological splits, metrics, and leakage-control rules.

    \item \textbf{A leakage-controlled multi-source data infrastructure.}
    We align 38,128 naturally caused FPA-FOD wildfire events with public discovery-time and static signals from FIRMS/VIIRS, gridMET, LANDFIRE, OpenStreetMap, and WorldPop, and provide model-ready tabular, temporal, spatial, and spatiotemporal representations.

    \item \textbf{A broad baseline suite across prediction paradigms.}
    We evaluate 16 representative models covering classical baselines and recent architectures for tabular, temporal, spatial, and spatiotemporal learning, establishing a reproducible performance reference for future IA prediction research.

    \item \textbf{Benchmark-driven insights into early wildfire risk.}
    Our experiments show that public discovery-time data provides meaningful but incomplete signal for IA failure prediction, identify FIRMS/VIIRS as the least redundant discovery-day source and fuel as the strongest static fallback source, and show that containment duration is weakly explained by discovery-time inputs.
\end{itemize}

\section{Benchmark Design and Data Construction}\label{sec:benchmark}

This section first defines the event-level IA failure prediction task and research questions, then describes the public data sources used in \benchmarkname{}, and finally presents the canonicalization pipeline that converts these sources into leakage-controlled, model-ready artifacts.

\subsection{Event-Level Task Design and Research Questions}
\label{sec:problem_formulation}

This subsection defines the event unit, input rule, label construction, and research questions for \benchmarkname{}. Let $i \in \{1,\ldots,n\}$ index wildfire events, and let $t_i^d$ denote the reported discovery time of event $i$. 
Let $a_i$ denote its final burned area, which is used only for label construction and never used as a model input. 
Let $\mathcal{C}$ denote the set of public source groups and $\mathcal{S} \subseteq \mathcal{C}$ denote any source subset. 
For event $i$, let $\boldsymbol{x}_i(\mathcal{S})$ denote the model-ready input constructed from sources in $\mathcal{S}$ using only public information available at or before $t_i^d$. 
The input $\boldsymbol{x}_i(\mathcal{S})$ may be tabular, temporal, spatial, or spatiotemporal depending on the model family. Following prior wildfire-management studies~\cite{korkola2024ia,xu2023initialattack,butler2025victoria,cardil2025suppression,arienti2006empirical}, we define the initial attack below.

\begin{definition}[Initial Attack]\label{def:initial_attack}
For a reported wildfire event $i$, the \emph{initial attack} phase is the first suppression phase beginning at $t_i^d$, during which first-arriving firefighting resources attempt to halt or contain fire spread.
A wildfire \emph{escapes initial attack} if this first response fails and the incident requires suppression beyond the initial response capacity.
\end{definition}

Because direct operational IA outcomes are not consistently reported in public national records, we follow prior size-based practice and define fires no larger than $10$\,ha as successful IA ($y_i=0$), fires at least $50$\,ha as IA failure ($y_i=1$), and exclude intermediate-size fires to reduce label ambiguity~\cite{xu2023initialattack,korkola2024ia,cardil2025suppression}. IA Failure prediction is then defined as:

\begin{problem}[IA Failure Prediction]\label{prob:ia_prediction}
Given the discovery-time input $\boldsymbol{x}_i(\mathcal{S})$ and binary IA label $y_i$, the task is to learn a predictor $p_i(\mathcal{S}) = f_{\boldsymbol{\theta}}\bigl(\boldsymbol{x}_i(\mathcal{S})\bigr)
\approx \Pr(y_i = 1 \mid \boldsymbol{x}_i(\mathcal{S}))$, where $p_i(\mathcal{S}) \in [0,1]$ is the predicted probability that wildfire event $i$ escapes initial attack. 
The full benchmark dataset is $\mathcal{D} = \{(\boldsymbol{x}_i(\mathcal{C}), y_i)\}_{i=1}^{n}$. In \benchmarkname{}, $\mathcal{C}$ includes FPA-FOD metadata, FIRMS/VIIRS fire-signal features, gridMET weather and fire-danger features, LANDFIRE vegetation, fuel, and topography features, OpenStreetMap access features, and WorldPop population-context features. 
Different choices of $\mathcal{S}$ instantiate the full-source, source-ablation, static-source, and weather-history tasks evaluated in later sections. Based on this task design, \benchmarkname{} studies five research questions.
\end{problem} 

\noindent \textbf{RQ1: Full-source predictability.}
How accurately can public discovery-time sources predict IA failure, and which input representation is most effective?

\noindent \textbf{RQ2: Source necessity.}
Under the full-input setting, which source groups contribute non-redundant predictive signal, and which become largely redundant once other sources are available?

\noindent \textbf{RQ3: Static-source value.}
When weather history and discovery-day satellite detections are unavailable, which static sources improve prediction beyond FPA-FOD metadata?

\noindent \textbf{RQ4: Weather-history sufficiency.}
How much near-discovery weather and fire-danger history is needed for IA failure prediction, and does a multi-day history improve over discovery-day conditions?

\noindent \textbf{RQ5: Containment-duration signal.}
Can the same discovery-time inputs predict containment duration, or is this post-discovery outcome driven by factors outside the benchmark input contract?

\begin{table*}[t]
\centering
\scriptsize
\caption{Dataset-source inventory for \benchmarkname. Each row corresponds to a downloaded public data product or raster layer. ``Canonical size'' reports the scale after preprocessing into event-level, daily, or patch-level benchmark data. Area is reported in $km^2$ for the nominal source coverage.}
\setlength{\tabcolsep}{2pt}
\renewcommand{\arraystretch}{1.08}
\label{tab:data_taxonomy}
\resizebox{\textwidth}{!}{%
\begin{tabular}{lllllllll}
\toprule
\textbf{Name} & \textbf{Source} & \textbf{Country} & \makecell[l]{\textbf{Area ($km^2$)}} & \makecell[l]{\textbf{Task}} & \textbf{Period} & \makecell[l]{\textbf{Spatial}\\\textbf{resolution}} & \makecell[l]{\textbf{Update}\\\textbf{frequency}} \\
\midrule
FPA-FOD & fire reports & USA & 9,834,000 & labels and metadata & 1992--2020 & event point & periodic release \\
FIRMS/VIIRS & fire signal & Worldwide & 149,000,000 & discovery-day thermal signal & 2012--2026 & 375 m & near-real-time \\
gridMET & weather & USA & 9,834,000 & weather and fire danger & 1979--2020 & $\sim$4 km & daily \\
Landfire & fuel & USA & 9,834,000 & Fuel Disturbance, Vegetation, Canopy & 2001--2024 & 30 m & versioned release \\
Landfire & vegetation & USA & 9,834,000 & Existing Vegetation & 2001--2024 & 30 m & versioned release \\
Landfire & topography & USA & 9,834,000 & Elevation, Aspect, Slope Degrees & 2001--2024 & 30 m & versioned release \\
OpenStreetMap & access & Worldwide & 149,000,000 & roads density and fire stations & 2004--2026 & vector & continuously updated \\
WorldPop & population & Worldwide & 149,000,000 & population density & 2000--2026 & $\sim$100 m & annual \\
\bottomrule
\end{tabular}%
}
\vspace{-2em}
\end{table*}

\begin{figure*}[t]
\centering
\includegraphics[width=1\textwidth]{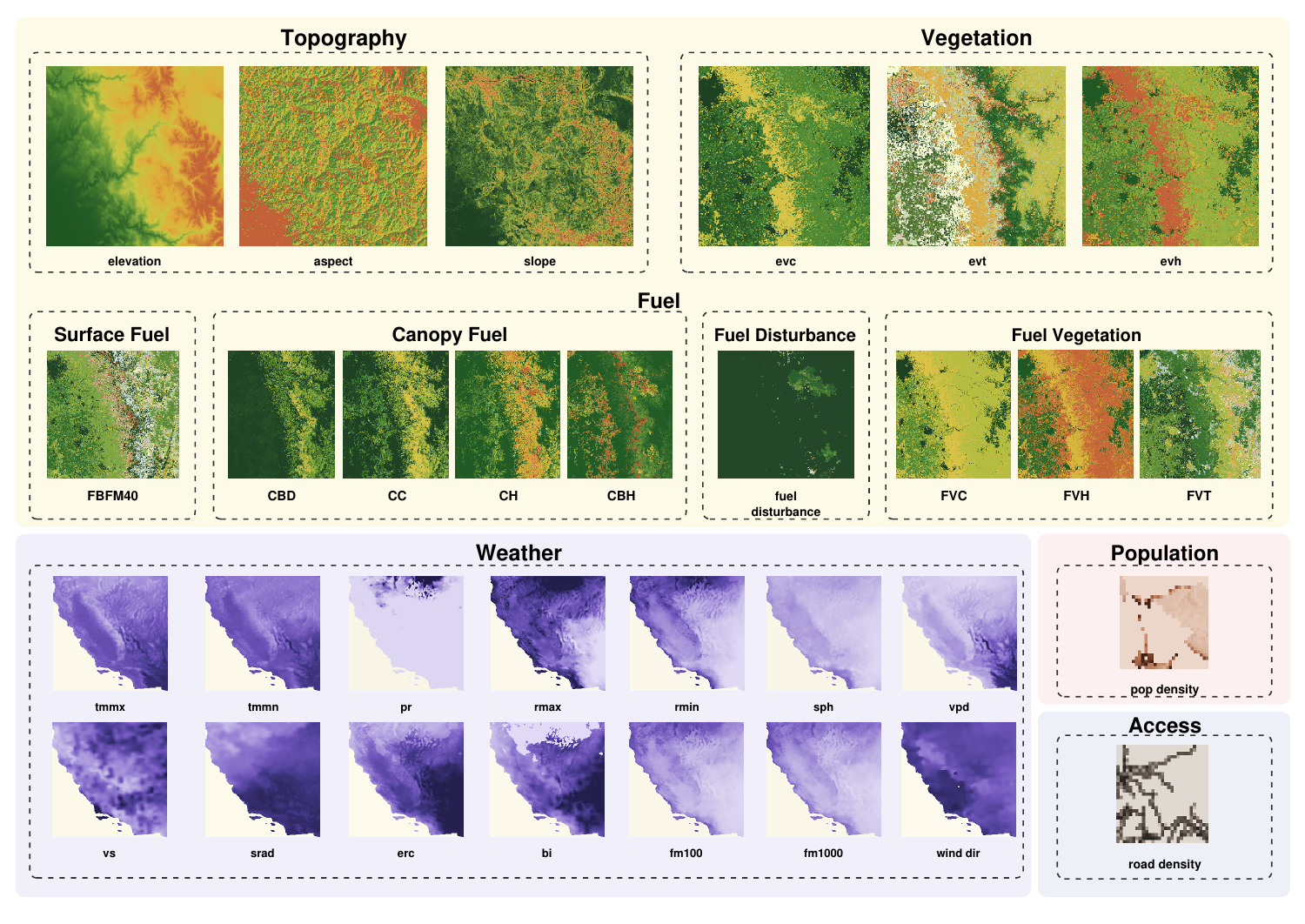}
\caption{Examples of event-centered source layers used by \benchmarkname. The visualization shows aligned topography, vegetation, fuel, weather, population, and access channels around a reported wildfire location. These layers are discovery-time or static inputs, not prediction targets.}
\label{fig:dataset_example}
\vspace{-2em}
\end{figure*}

\vspace{-0.5em}
\subsection{Data Sources and Collection}
\label{sec:data_sources_collection}
\vspace{-0.5em}
Table~\ref{tab:data_taxonomy} summarizes the public data products used in \benchmarkname{}, covering the event backbone, discovery-day fire signal, weather and fire-danger history, static landscape context, access proxies, and population context. Figure~\ref{fig:dataset_example} illustrates how these event-centered layers are aligned around each reported wildfire before being converted into benchmark features.

\noindent\textbf{Event backbone.}
\benchmarkname{} uses FPA-FOD as the event backbone~\cite{short2017fpafod}, retaining 38,128 naturally caused wildfires from 2016--2020 in the contiguous United States with valid identifiers, discovery dates, coordinates, and positive final sizes. 
This yields a rare-event task, with IA failure rates of 7.55\%, 6.07\%, and 8.32\% in train, validation, and test.

\noindent\textbf{Discovery-day fire signal.}
FIRMS/VIIRS provides 375 m active-fire observations used only as discovery-day thermal evidence, not as the event definition~\cite{nasa_firms,schroeder2014viirs}. 
We match 1,738 detections to nearby FPA-FOD events and summarize count, fire radiative power, brightness, and patch-level channels; unmatched events receive zero-valued thermal features.

\noindent\textbf{Weather.}
gridMET provides daily meteorological and fire-danger variables through discovery day at approximately 4 km resolution~\cite{abatzoglou2013gridmet}. For each event we extract a D-4:D discovery-window sequence covering temperature, humidity, wind, energy release component, burning index, and fuel moisture.

\noindent\textbf{Topography.}
LANDFIRE topography provides CONUS-scale terrain layers at 30 m resolution~\cite{landfire}. 
We use elevation, slope, and aspect as fixed terrain attributes around each reported fire location.

\noindent\textbf{Vegetation.}
LANDFIRE vegetation provides 30 m CONUS-wide landscape structure and plant-community information~\cite{landfire}. 
We use Existing Vegetation Type (EVT), Existing Vegetation Cover (EVC), and Existing Vegetation Height (EVH) to describe local vegetation composition and structure.

\noindent\textbf{Fuel.}
LANDFIRE fuel products provide 30 m combustible-material layers over the contiguous United States~\cite{landfire}. 
We use surface fuel (FBFM40), canopy fuel variables, fuel disturbance, and fuel vegetation layers to represent persistent fuel conditions relevant to fire growth potential.

\noindent\textbf{Access.}
OpenStreetMap provides access proxies, including drivable-road density and fire-station proximity, at both event and patch levels~\cite{osm}. 
The benchmark uses a 2020 extract with 26.0M drivable-road geometries, 12.8M km of projected road linework, and 27,989 fire-station geometries.

\noindent\textbf{Population.}
WorldPop provides prior-year annual population density for each fire year, using the preceding year's raster to avoid future information~\cite{worldpop}. The annual rasters use 0.000833$^\circ$ cells, approximately 100 m resolution. The 2019 raster contains 430,711 $\times$ 62,976 cells.

\subsection{Canonicalization Pipeline}
\label{sec:canonicalization_pipeline}

\begin{figure*}[t]
\centering
\includegraphics[width=1\textwidth]{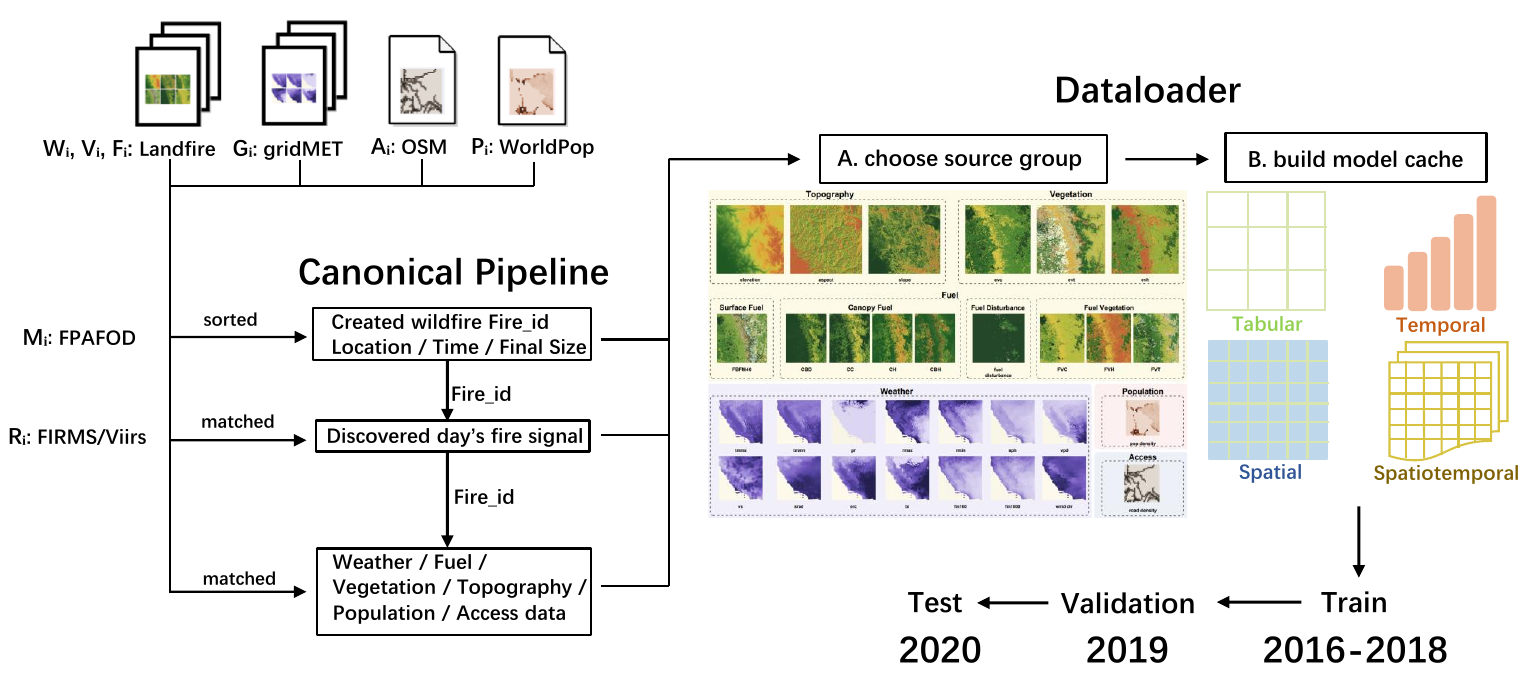}
\caption{
The full construction pipeline aligns raw data into canonicalized data, loads it into a training-ready cache, and defines the train/validation/test setting.
}
\label{fig:pipeline2}
\vspace{-1.5em}
\end{figure*}

Figure~\ref{fig:pipeline2} summarizes the construction pipeline used to enforce a common event unit across all sources. 
Each retained FPA-FOD record defines one benchmark sample with a unique \texttt{fire\_id}; its discovery date, location, and fire year determine how external public sources are attached. 
Raw sources are aligned to these event records and canonicalized into event-level and patch-level artifacts keyed by \texttt{fire\_id}. 
The dataloader then applies the requested source setting and discovery-time rules before converting the canonical artifacts into tabular, temporal, spatial, and spatiotemporal model caches.

\noindent\textbf{Label construction.}
Each retained FPA-FOD record defines one benchmark sample with a unique fire\_id. The event's discovery date, location, and fire year serve as the join key for attaching all external sources. Final fire size and containment timestamps are used solely for label construction and are excluded from model inputs via a forbidden-feature manifest.

\noindent\textbf{Fire signal matching.}
FIRMS/VIIRS detections are attached to FPA-FOD events only on the reported discovery day. 
Each VIIRS detection is assigned to at most one nearby FPA-FOD event within the matching radius. 
Unmatched VIIRS detections are discarded, while FPA-FOD events without matched VIIRS detections are retained with zero-valued thermal features and a detection indicator. 
D+1 and later detections are excluded to avoid post-discovery fire-growth leakage.

\noindent\textbf{Environmental and contextual features.}
We extract gridMET weather and fire-danger features through discovery day, and use LANDFIRE fuel, vegetation, and topography as static landscape context. OpenStreetMap provides access and response-proximity features, while prior-year WorldPop rasters provide population context. All features are stored as \texttt{fire\_id}-indexed canonical tables.

\noindent\textbf{Spatial artifacts and outputs.}
For spatial models, we build an event-centered \(29\times29\) grid at 375 m resolution in EPSG:5070 around each discovery location. Raster, vector, and point sources are sampled on this grid to produce spatial and spatiotemporal caches. The canonical outputs include event tables, source-specific features, daily weather records, spatial artifacts, and a master table with fixed splits, input windows, and forbidden-feature rules.

\vspace{-0.5em}
\section{Tasks, Metrics \& Evaluation Protocol}\label{sec:tasks}
\vspace{-0.5em}
In this section we introduce how \benchmarkname{} evaluates IA failure prediction. 
All tasks share the same event unit, label rule, discovery-time input constraint, and leakage-control policy, while varying the source subset, weather window, representation family, or prediction target. 
We then specify metrics for rare-event classification and auxiliary containment-duration prediction.

\subsection{Evaluation Tasks}
\label{sec:evaluation_contracts}
All tasks use the label $y_i$, source set $\mathcal{C}$, and input notation $\boldsymbol{x}_i(\mathcal{S})$ from Problem~\ref{prob:ia_prediction}, and all inputs follow the discovery-time leakage-control rule in Section~\ref{sec:problem_formulation}.

\noindent\textbf{Task 1: Full-source predictability.}
This task evaluates whether the complete public discovery-time source stack can predict initial attack failure.
Each model receives $\boldsymbol{x}_i(\mathcal{C})$ and predicts
\[
p_i^{\mathrm{IA}} = f_{\boldsymbol{\theta}}\bigl(\boldsymbol{x}_i(\mathcal{C})\bigr),
\]
where $p_i^{\mathrm{IA}} \in [0,1]$ is the predicted probability that event $i$ escapes initial attack.

\noindent\textbf{Task 2: Source necessity under full input.}
This task measures whether each source group contributes non-redundant information when all other sources are available.
For each source group $\mathcal{Q} \in \mathcal{C} \setminus \{\mathcal{M}\}$, the model receives $\boldsymbol{x}_i(\mathcal{C} \setminus \{\mathcal{Q}\})$ and predicts $y_i$.
A large performance drop after removing $\mathcal{Q}$ indicates that the removed source contains information not recovered by the remaining source groups.

\noindent\textbf{Task 3: Static-source value without dynamic observations.}
This task evaluates which static source remains useful when dynamic observations are unavailable.
We remove FIRMS/VIIRS and weather by excluding $\mathcal{R}$ and $\mathcal{W}$, and compare metadata plus one static source group, $\boldsymbol{x}_i(\{\mathcal{M}, \mathcal{Q}\})$, for each $\mathcal{Q} \in \{\mathcal{V}, \mathcal{F}, \mathcal{T}, \mathcal{A}, \mathcal{P}\}$.
We also evaluate $\boldsymbol{x}_i(\{\mathcal{M}\})$ as the metadata-only lower bound.

\noindent\textbf{Task 4: Weather-history sufficiency.}
This task evaluates how much retrospective weather history is needed for initial attack failure prediction.
Let $\ell \in \{1,2,3,4,5\}$ denote the number of days in the weather window ending on discovery day $t_i^{d}$, and let $\mathcal{W}^{(\ell)} \subseteq \mathcal{W}$ denote the $\ell$-day gridMET weather and fire-danger history.
Each model receives $\boldsymbol{x}_i(\{\mathcal{M}, \mathcal{W}^{(\ell)}\})$ and predicts $y_i$.
Comparing performance across values of $\ell$ tests whether longer weather histories improve prediction beyond weather near discovery time.

\noindent\textbf{Task 5: Auxiliary containment-duration prediction.}
This task tests whether the same discovery-time public inputs can predict containment duration.
For events with valid timestamps, let $t_i^{c}$ denote containment time; we define containment duration $h_i = t_i^{c} - t_i^{d}$ and log-transformed target $y_i^{\mathrm{TTC}} = \log(1 + h_i)$.
The model receives $\boldsymbol{x}_i(\mathcal{C})$ and predicts
\[
\hat{y}_i^{\mathrm{TTC}} = g_{\boldsymbol{\phi}}\bigl(\boldsymbol{x}_i(\mathcal{C})\bigr),
\qquad
\hat{h}_i = \exp(\hat{y}_i^{\mathrm{TTC}}) - 1,
\]
where $g_{\boldsymbol{\phi}}$ is a predictor parameterized by vector $\boldsymbol{\phi}$.
This auxiliary task is not used to define initial attack failure.
It serves as a stress test of how much post-discovery suppression outcome can be explained by discovery-time public data.

\subsection{Evaluation Metrics}
\label{sec:metrics}

\noindent\textbf{IA failure metrics.}
For IA failure prediction, we use \textbf{AUPRC} as the primary metric because escaped fires are rare and the benchmark emphasizes ranking high-risk events near the top.
We also report \textbf{AUROC}, \textbf{Recall@5\%}, \textbf{F1}, \textbf{Brier score}, and \textbf{ECE}.
These metrics measure ranking quality, early-warning recall, thresholded classification performance, probability error, and calibration.

\noindent\textbf{Containment-duration metrics.}
For auxiliary containment-duration prediction, we use \textbf{MAE} in hours as the primary metric after converting log-space predictions back to containment hours.
We also report \textbf{RMSE}, \textbf{MedianAE}, \textbf{log-space MAE}, $R^2$, and \textbf{Spearman correlation}.
These metrics measure large-error sensitivity, typical-event error, training-scale error, explained variance, and ranking consistency across containment durations.
\section{Experimental Setup}
\label{sec:baseline}

All experiments use the same canonical \benchmarkname{} artifacts. 
Each sample is one FPA-FOD natural wildfire event indexed by \texttt{fire\_id}. 
We use a chronological split, with 2016--2018 fires for training, 2019 for validation, and 2020 for testing. 
Across all experiments, we keep the event set, labels, forbidden-feature list, and five-seed protocol fixed, so that differences reflect source choices, representations, or model families rather than changes in benchmark construction.

\subsection{Model-Ready Representations}
\label{sec:model_ready_representations}

\benchmarkname{} compares model families under the same event-level IA failure prediction contract. 
After public sources are aligned by \texttt{fire\_id}, the dataloader applies the requested task and source setting, removes forbidden fields, fits preprocessing statistics on the training split only, and exports model-ready caches. 
This ensures that models use the same events, labels, splits, and leakage policy, while differing only in input representation. 
As summarized in Table~\ref{tab:model_suite}, \benchmarkname{} provides four views: a tabular view that aggregates each event into structured features, a temporal view that preserves the \(D{-}4{:}D\) weather and fire-danger history, a spatial view that represents each fire as an event-centered \(29\times29\) patch, and a spatiotemporal view that combines patch-level context with short weather dynamics. 
All views produce the same scalar event-level output, either IA failure probability or auxiliary containment duration, rather than a fire mask, perimeter, or spread trajectory.

\begin{table*}[t]
\centering
\scriptsize
\caption{Model-ready representations and baseline families in \benchmarkname{}.}
\label{tab:model_suite}
\begin{tabular*}{\textwidth}{@{\extracolsep{\fill}}L{0.12\textwidth}L{0.25\textwidth}L{0.66\textwidth}@{}}
\toprule
\textbf{Family} & \textbf{Input shape} & \textbf{Models} \\
\midrule
Tabular & $N \times F$ & logistic regression, XGBoost, MLP \\
Temporal & $N \times T \times F_t$ plus $N \times F_s$ & GRU, TCN, Transformer \\
Spatial & $N \times C \times 29 \times 29$ & ResNet18-UNet, ResNet50-UNet, Swin-UNet, SegFormer \\
Spatiotemporal & $N \times T \times C \times 29 \times 29$ & ConvLSTM, ConvGRU, ResNet3D, PredRNN-V2, UTAE, SwinLSTM \\
\bottomrule
\end{tabular*}
\end{table*}

\subsection{Baseline Model Suite}
\label{sec:baseline_models}

We evaluate 16 baselines across tabular, temporal, spatial, and spatiotemporal families. 
The suite covers structured-data models, recurrent and attention-based sequence encoders, convolutional and transformer-based spatial encoders, and recurrent or 3D spatiotemporal architectures~\cite{chen2016xgboost,cho2014learning,bai2018empirical,vaswani2017attention,he2016deep,ronneberger2015unet,liu2021swin,cao2021swinunet,xie2021segformer,shi2015convlstm,wang2022predrnn,garnot2021utae,tang2023swinlstm}. 
Segmentation-style architectures are adapted by replacing dense prediction heads with global pooling and a scalar output, so every model predicts either initial attack failure probability or auxiliary containment duration.

\subsection{Training and Leakage Control}
\label{sec:training_preprocessing}

All preprocessing is fit on the training split only and then fixed for validation and test events. 
Before training, the dataloader removes forbidden fields, including final fire size, containment timestamps, MTBS identifiers, post-discovery detections, and label-derived variables. 
Discovery-day FIRMS/VIIRS features are allowed, but later detections are excluded; unmatched events are retained with zero-valued thermal features and a detection indicator.

\section{Experiments}\label{sec:result}
\newcommand{\pmstd}[1]{${}_{\scriptstyle \pm #1}$}

We organize the experiments around the five research questions defined in Section~\ref{sec:problem_formulation}. 
All experiments share the same event unit, chronological split, leakage policy, and evaluation protocol, making results comparable across model families, source combinations, weather windows, and prediction targets. 
\begin{table*}[t]
\centering
\small
\caption{Full-Source Predictability leaderboard. Values are percentages and shown as mean$\pm$std. AUPRC is the primary metric; AUROC and Recall@5\% evaluate ranking, F1 evaluates a validation-selected operating point, and Brier/ECE evaluate calibration. Best values are bolded.}
\label{tab:task1_full_leaderboard}
\resizebox{\textwidth}{!}{%
\begin{tabular}{llcccccc}
\toprule
\textbf{Family} & \textbf{Model} & \textbf{AUPRC}$\uparrow$ & \textbf{AUROC}$\uparrow$ & \textbf{Recall@5\%}$\uparrow$ & \textbf{F1}$\uparrow$ & \textbf{Brier}$\downarrow$ & \textbf{ECE}$\downarrow$ \\
\midrule
\multirow{3}{*}{Tabular} & LogisticRegression & 43.9\pmstd{0.0} & 81.5\pmstd{0.0} & 32.9\pmstd{0.0} & 42.5\pmstd{0.0} & 13.6\pmstd{0.0} & 15.8\pmstd{0.0} \\
 & XGBoost & \textbf{53.3\pmstd{0.3}} & \textbf{87.1\pmstd{0.1}} & \textbf{38.0\pmstd{0.6}} & \textbf{49.5\pmstd{0.6}} & 9.1\pmstd{0.1} & 13.5\pmstd{0.1} \\
 & MLP & 50.5\pmstd{0.5} & 83.4\pmstd{0.5} & 36.7\pmstd{0.3} & 47.1\pmstd{0.6} & 6.1\pmstd{0.1} & 5.9\pmstd{0.7} \\
\midrule
\multirow{3}{*}{Temporal} & GRU & 49.5\pmstd{0.1} & 82.9\pmstd{0.4} & 36.3\pmstd{0.6} & 46.0\pmstd{0.8} & 5.7\pmstd{0.0} & \textbf{0.9\pmstd{0.3}} \\
 & TCN & 50.4\pmstd{0.3} & 83.7\pmstd{0.5} & 36.4\pmstd{0.3} & 46.7\pmstd{0.6} & 5.8\pmstd{0.0} & 3.5\pmstd{0.6} \\
 & Transformer & 50.6\pmstd{0.3} & 84.0\pmstd{0.4} & 36.5\pmstd{0.3} & 46.8\pmstd{0.6} & 5.6\pmstd{0.0} & 0.9\pmstd{0.4} \\
\midrule
\multirow{4}{*}{Spatial} & ResNet18-UNet & 50.2\pmstd{1.7} & 84.8\pmstd{0.7} & 35.4\pmstd{1.1} & 47.2\pmstd{0.8} & 5.8\pmstd{0.3} & 4.3\pmstd{1.4} \\
 & ResNet50-UNet & 49.4\pmstd{0.6} & 84.0\pmstd{0.9} & 34.9\pmstd{1.2} & 44.8\pmstd{1.5} & 5.8\pmstd{0.1} & 3.3\pmstd{0.3} \\
 & Swin-UNet & 51.1\pmstd{1.2} & 85.3\pmstd{0.6} & 36.5\pmstd{1.8} & 46.7\pmstd{1.9} & 5.6\pmstd{0.1} & 2.6\pmstd{0.6} \\
 & SegFormer & 50.6\pmstd{0.5} & 84.9\pmstd{0.9} & 35.8\pmstd{0.4} & 46.6\pmstd{1.2} & 5.7\pmstd{0.1} & 2.3\pmstd{0.3} \\
\midrule
\multirow{6}{*}{Spatiotemporal} & ConvLSTM & 38.9\pmstd{1.7} & 83.8\pmstd{0.3} & 29.7\pmstd{2.0} & 40.4\pmstd{1.0} & 6.3\pmstd{0.1} & 3.3\pmstd{0.4} \\
 & ConvGRU & 41.2\pmstd{1.6} & 84.1\pmstd{0.6} & 31.6\pmstd{0.9} & 41.6\pmstd{1.1} & 6.2\pmstd{0.1} & 3.5\pmstd{0.2} \\
 & ResNet3D & 51.3\pmstd{1.7} & 84.5\pmstd{2.0} & 36.5\pmstd{1.5} & 46.7\pmstd{1.1} & 5.7\pmstd{0.1} & 3.5\pmstd{1.3} \\
 & PredRNN-V2 & 39.6\pmstd{1.6} & 83.4\pmstd{0.4} & 30.8\pmstd{1.1} & 41.8\pmstd{2.0} & 6.3\pmstd{0.1} & 2.7\pmstd{0.5} \\
 & UTAE & 51.3\pmstd{1.5} & 85.1\pmstd{1.2} & 36.2\pmstd{0.8} & 46.6\pmstd{1.1} & 5.7\pmstd{0.2} & 3.3\pmstd{1.2} \\
 & SwinLSTM & 51.7\pmstd{1.2} & 85.7\pmstd{0.9} & 37.3\pmstd{1.3} & 47.1\pmstd{2.3} & \textbf{5.6\pmstd{0.1}} & 2.4\pmstd{0.7} \\
\bottomrule
\end{tabular}%
}
\end{table*}

\begin{figure*}[t]
\centering
\includegraphics[width=1\textwidth]{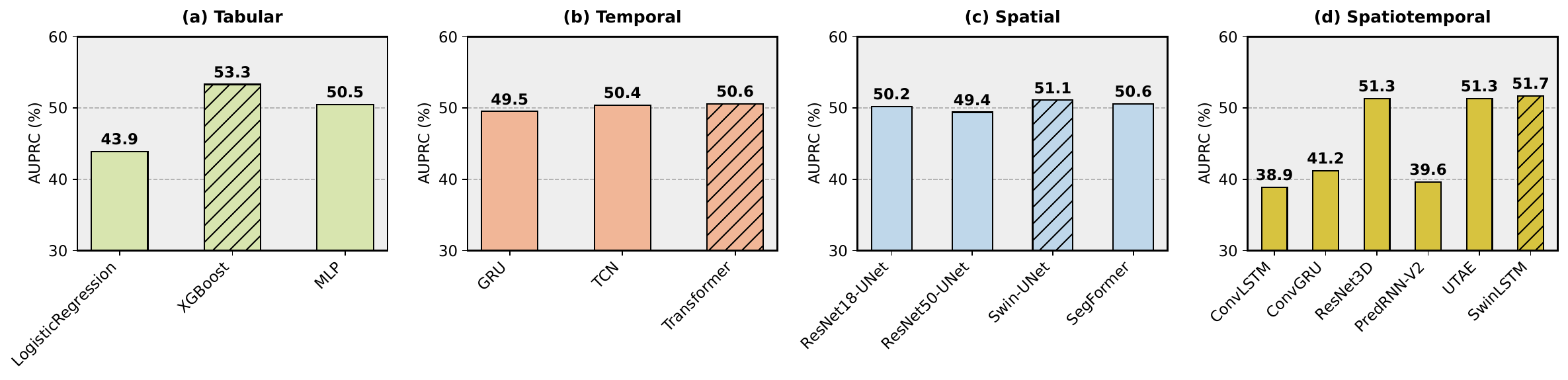}
\vspace{-2em}
\caption{Full-Source Initial Attack Predictability leaderboard.
}
\label{fig:full}
\end{figure*}

\subsection{Full-Source Predictability}

To answer \textbf{RQ1}, we evaluate the full-source setting using all discovery-time environmental and contextual inputs across tabular, temporal, spatial, and spatiotemporal model families. 
Table~\ref{tab:task1_full_leaderboard} reports the full metric suite over five random seeds, and Figure~\ref{fig:full} visualizes the AUPRC leaderboard by representation family. We find that:
(1) The full-source setting contains a clear predictive signal: all major representation families substantially outperform a trivial rare-event baseline, showing that initial attack failure can be predicted from public discovery-time and static contextual variables. 
(2) The task behaves primarily as event-level risk prediction rather than dense spatial forecasting. 
Higher-capacity spatial and spatiotemporal models do not automatically outperform structured event-level models, and the best results are obtained by XGBoost. 
(3) Patch-based models are still competitive, indicating that local landscape and discovery-day spatial context are useful, but they do not replace aggregated event-level summaries. 
At the same time, the best Recall@5\% remains limited, showing that public discovery-time inputs do not fully determine suppression outcome.

\subsection{Source Necessity under Full Input.}

\begin{table}[t]
\centering
\small
\setlength{\tabcolsep}{2pt}
\renewcommand{\arraystretch}{0.90}
\caption{Source Necessity under Full Input for initial attack escape prediction. Each cell is AUPRC in percent after removing one source from the full input, reported as mean$\pm$standard deviation over five random seeds. The final column reports the unablated full-input result.}
\label{tab:without_source_ablation}
\begin{tabular*}{\linewidth}{@{\extracolsep{\fill}}lcccccccc@{}}
\toprule
\textbf{Model}
& \makecell{\textbf{w/o}\\[-3pt]\textbf{FIRMS}}
& \makecell{\textbf{w/o}\\[-3pt]\textbf{Weather}}
& \makecell{\textbf{w/o}\\[-3pt]\textbf{Vegetation}}
& \makecell{\textbf{w/o}\\[-3pt]\textbf{Fuel}}
& \makecell{\textbf{w/o}\\[-3pt]\textbf{Topography}}
& \makecell{\textbf{w/o}\\[-3pt]\textbf{Access}}
& \makecell{\textbf{w/o}\\[-3pt]\textbf{Population}}
& \textbf{Full} \\
\midrule
XGBoost & 38.3\pmstd{0.5} & 50.5\pmstd{0.5} & \textbf{53.6\pmstd{0.3}} & 52.1\pmstd{0.3} & 53.1\pmstd{0.2} & 53.6\pmstd{0.4} & 51.2\pmstd{0.4} & 53.3\pmstd{0.3} \\
Transformer & 32.3\pmstd{0.2} & 49.2\pmstd{0.3} & \textbf{51.4\pmstd{0.3}} & 49.7\pmstd{0.3} & 50.7\pmstd{0.2} & 50.2\pmstd{0.5} & 50.3\pmstd{0.6} & 50.6\pmstd{0.3} \\
Swin-UNet & 30.4\pmstd{1.0} & 48.8\pmstd{1.2} & 48.2\pmstd{1.5} & 49.5\pmstd{1.1} & 50.6\pmstd{0.9} & 46.2\pmstd{1.1} & 50.0\pmstd{1.4} & \textbf{51.1\pmstd{1.2}} \\
SwinLSTM & 33.4\pmstd{1.1} & 48.1\pmstd{0.8} & 51.3\pmstd{0.7} & \textbf{52.2\pmstd{0.4}} & 51.0\pmstd{1.0} & 50.6\pmstd{0.9} & 51.6\pmstd{1.1} & 51.7\pmstd{1.2} \\
\bottomrule
\end{tabular*}
\end{table}
To answer \textbf{RQ2}, we perform leave-one-source-out ablations and compare each setting with the full-source protocol. 
This measures marginal necessity: a large AUPRC drop indicates that the removed source provides signal not recovered by the remaining inputs. 
Table~\ref{tab:without_source_ablation} shows that FIRMS/VIIRS is the least redundant source, as removing discovery-day thermal evidence causes the largest degradation across representative models. 
Weather provides a smaller but consistent gain, suggesting partial overlap with other contextual variables. 
Vegetation, fuel, topography, access, and population have marginal or even slightly negative effects in the full-input setting, indicating that their information is partly covered once FIRMS/VIIRS, weather, metadata, and other contextual sources are available.

\subsection{Static-source value without dynamic observations.}

\begin{wraptable}[7]{r}{0.55\textwidth}
\vspace{-2em}
\centering
\tiny
\setlength{\tabcolsep}{1.6pt}
\renewcommand{\arraystretch}{0.78}
\caption{Static-source ablation. AUPRC is reported in percentage over five seeds.}
\label{tab:source_only_ablation}
\resizebox{\linewidth}{!}{
\begin{tabular}{lcccccc}
\toprule
\textbf{Model} & \textbf{Vegetation} & \textbf{Fuel} & \textbf{Topography} & \textbf{Access} & \textbf{Population} & \textbf{FPA-FOD} \\
\midrule
XGBoost & 25.6\pmstd{0.2} & \textbf{30.3\pmstd{0.6}} & 25.6\pmstd{0.5} & 24.2\pmstd{0.3} & 26.7\pmstd{0.4} & 23.6\pmstd{0.5} \\
Transformer & 25.3\pmstd{0.6} & \textbf{28.8\pmstd{0.3}} & 24.6\pmstd{0.2} & 23.9\pmstd{0.5} & 20.9\pmstd{0.5} & 22.8\pmstd{0.4} \\
Swin-UNet & 22.1\pmstd{2.4} & \textbf{23.3\pmstd{1.7}} & 21.2\pmstd{3.0} & 20.0\pmstd{1.6} & 17.0\pmstd{0.8} & 15.9\pmstd{0.7} \\
SwinLSTM & 26.4\pmstd{1.2} & \textbf{26.7\pmstd{2.1}} & 21.1\pmstd{0.9} & 20.8\pmstd{1.2} & 15.5\pmstd{1.0} & 14.5\pmstd{1.3} \\
\bottomrule
\end{tabular}
}
\vspace{-2em}
\end{wraptable}

To answer \textbf{RQ3}, we remove both dynamic sources and evaluate two conditions: FPA-FOD metadata alone as the lower bound, and FPA-FOD metadata combined with one static source group at a time. This fallback setting measures which static information remains useful without weather history or discovery-day satellite evidence. Table~\ref{tab:source_only_ablation} shows that fuel is the strongest static source across representative models, suggesting that combustible landscape structure provides meaningful background risk signal. The metadata-only condition confirms the lower bound: all static sources provide a lift over using metadata alone, with fuel showing the largest and most consistent gain. However, static sources remain insufficient on their own: vegetation, topography, access, and population are weaker and more model-dependent, indicating that static context can support early risk assessment but cannot replace current fire signal or near-discovery weather observations.

\begin{table*}[t]
\centering
\small
\caption{Containment-Duration prediction over five random seeds. Models predict log containment hours; MAE, RMSE, and MedianAE are reported after converting predictions back to hours. Lower is better for error metrics, and higher is better for $R^2$ and Spearman correlation.}
\label{tab:task2_containment_leaderboard}
\resizebox{\textwidth}{!}{%
\begin{tabular}{llcccccc}
\toprule
\textbf{Family} & \textbf{Model} & \textbf{MAE h}$\downarrow$ & \textbf{RMSE h}$\downarrow$ & \textbf{MedianAE h}$\downarrow$ & \textbf{log MAE}$\downarrow$ & \textbf{$R^2$}$\uparrow$ & \textbf{Spearman}$\uparrow$ \\
\midrule
\multirow{3}{*}{Tabular} & RidgeRegression & 39.3\pmstd{0.0} & 129.1\pmstd{0.0} & 8.4\pmstd{0.0} & 1.319\pmstd{0.000} & 0.058\pmstd{0.000} & 0.345\pmstd{0.000} \\
 & XGBoost & \textbf{35.1\pmstd{0.0}} & \textbf{121.3\pmstd{0.2}} & 8.2\pmstd{0.1} & \textbf{1.197\pmstd{0.002}} & 0.172\pmstd{0.004} & \textbf{0.411\pmstd{0.002}} \\
 & MLP & 35.5\pmstd{0.3} & 125.6\pmstd{0.8} & \textbf{4.9\pmstd{0.2}} & 1.209\pmstd{0.011} & 0.176\pmstd{0.016} & 0.396\pmstd{0.005} \\
\midrule
\multirow{3}{*}{Temporal} & GRU & 36.4\pmstd{0.3} & 126.2\pmstd{3.1} & 7.3\pmstd{0.1} & 1.222\pmstd{0.006} & \textbf{0.177\pmstd{0.011}} & 0.375\pmstd{0.007} \\
 & TCN & 36.1\pmstd{0.3} & 124.7\pmstd{0.7} & 6.6\pmstd{0.3} & 1.226\pmstd{0.003} & 0.157\pmstd{0.007} & 0.376\pmstd{0.005} \\
 & Transformer & 36.0\pmstd{0.3} & 124.1\pmstd{1.3} & 6.9\pmstd{0.5} & 1.216\pmstd{0.010} & 0.168\pmstd{0.015} & 0.377\pmstd{0.007} \\
\midrule
\multirow{4}{*}{Spatial} & ResNet18-UNet & 36.0\pmstd{0.3} & 125.2\pmstd{0.4} & 8.1\pmstd{0.1} & 1.231\pmstd{0.006} & 0.150\pmstd{0.013} & 0.365\pmstd{0.011} \\
 & ResNet50-UNet & 37.2\pmstd{2.3} & 128.5\pmstd{2.6} & 8.4\pmstd{0.5} & 1.263\pmstd{0.014} & 0.110\pmstd{0.018} & 0.338\pmstd{0.009} \\
 & Swin-UNet & 36.9\pmstd{0.4} & 128.7\pmstd{6.0} & 8.3\pmstd{0.5} & 1.274\pmstd{0.026} & 0.114\pmstd{0.028} & 0.333\pmstd{0.024} \\
 & SegFormer & 36.9\pmstd{0.6} & 126.9\pmstd{2.8} & 8.3\pmstd{0.4} & 1.263\pmstd{0.021} & 0.106\pmstd{0.025} & 0.337\pmstd{0.019} \\
\midrule
\multirow{6}{*}{Spatiotemporal} & ConvLSTM & 36.5\pmstd{0.1} & 125.8\pmstd{0.5} & 7.8\pmstd{0.4} & 1.251\pmstd{0.009} & 0.136\pmstd{0.013} & 0.341\pmstd{0.008} \\
 & ConvGRU & 36.4\pmstd{0.1} & 125.5\pmstd{0.4} & 7.8\pmstd{0.2} & 1.243\pmstd{0.004} & 0.150\pmstd{0.008} & 0.353\pmstd{0.005} \\
 & ResNet3D & 37.5\pmstd{2.3} & 127.4\pmstd{1.2} & 8.4\pmstd{0.6} & 1.266\pmstd{0.042} & 0.126\pmstd{0.055} & 0.356\pmstd{0.022} \\
 & PredRNN-V2 & 36.4\pmstd{0.1} & 125.6\pmstd{0.4} & 7.9\pmstd{0.3} & 1.243\pmstd{0.008} & 0.145\pmstd{0.009} & 0.353\pmstd{0.008} \\
 & UTAE & 37.6\pmstd{2.1} & 131.7\pmstd{3.8} & 6.6\pmstd{1.9} & 1.300\pmstd{0.060} & 0.080\pmstd{0.114} & 0.309\pmstd{0.019} \\
 & SwinLSTM & 36.3\pmstd{0.2} & 125.6\pmstd{0.7} & 7.8\pmstd{0.6} & 1.231\pmstd{0.014} & 0.151\pmstd{0.016} & 0.365\pmstd{0.010} \\
\bottomrule
\end{tabular}%
}
\end{table*}

\subsection{Weather-History Sufficiency.}

\begin{wraptable}{r}{0.53\textwidth}
\vspace{-2em}
\centering
\footnotesize
\setlength{\tabcolsep}{2pt}
\renewcommand{\arraystretch}{0.85}
\caption{Weather-history sufficiency for initial attack escape prediction. Each cell is AUPRC in percentage. Best value in each row is bolded.}
\label{tab:weather_days_ablation}
\begin{tabular}{lccccc}
\toprule
\textbf{Model} & \textbf{1d} & \textbf{2d} & \textbf{3d} & \textbf{4d} & \textbf{5d} \\
\midrule
XGBoost & \textbf{20.1\pmstd{0.2}} & 18.8\pmstd{0.3} & 19.0\pmstd{0.2} & 19.1\pmstd{0.2} & 19.0\pmstd{0.6} \\
Transformer & 25.1\pmstd{0.2} & \textbf{25.3\pmstd{0.3}} & 24.7\pmstd{0.6} & 24.2\pmstd{0.8} & 23.8\pmstd{0.8} \\
Swin-UNet & 22.4\pmstd{0.6} & \textbf{22.8\pmstd{1.0}} & 22.7\pmstd{1.2} & 21.8\pmstd{1.1} & 21.6\pmstd{2.7} \\
SwinLSTM & 23.1\pmstd{1.3} & \textbf{23.7\pmstd{0.6}} & 22.7\pmstd{1.7} & 23.1\pmstd{1.4} & 23.0\pmstd{1.4} \\
\bottomrule
\end{tabular}
\vspace{-1.2em}
\end{wraptable}

To answer \textbf{RQ4}, we vary the weather and fire-danger window from discovery day to the full \(D{-}4{:}D\) history while keeping the initial attack failure target fixed. Table~\ref{tab:weather_days_ablation} shows that longer weather histories do not consistently improve prediction: the best or near-best AUPRC values usually occur with one or two days of weather, while adding three to five days provides little additional benefit. This suggests that, under the current event-level target and gridMET feature design, the most useful weather signal is concentrated near discovery time, even though longer antecedent conditions may remain physically relevant.

\subsection{Auxiliary Containment-Duration Signal.}

To answer \textbf{RQ5}, we test the boundary of the discovery-time benchmark contract by replacing the initial attack label with log containment hours. Table~\ref{tab:task2_containment_leaderboard} shows that XGBoost remains the strongest baseline, indicating that public discovery-time inputs contain some severity signal. However, the low \(R^2\) suggests that containment duration is only weakly explained by these inputs and is largely shaped by post-discovery resources, tactics, weather evolution, and reporting practices. This reinforces our main design choice: \benchmarkname{} is best suited for early event-level triage, while containment duration serves as an auxiliary stress test of the benchmark's limits.

\section{Conclusion}
\label{sec:conclusion}

We introduce \benchmarkname{}, a reproducible AI benchmark for event-level wildfire initial attack failure prediction from public U.S. national environmental data. \benchmarkname{} fixes the event unit, discovery-time input contract, leakage policy, temporal split, metrics, and model-ready representations, enabling fair comparison across tabular, temporal, spatial, and spatiotemporal models. Experiments show that discovery-time public data contains useful early-risk signals, while also revealing source-specific contributions and the limits of containment-duration prediction.

\bibliographystyle{plain}
\bibliography{ref}

\appendix

\section*{Supplementary Materials}

\section{Metric Definitions}

\noindent\textbf{Task 1 classification metrics.}
For the initial attack failure task, let $p_i^{\mathrm{IA}}$ denote the predicted failure probability and $y_i^{\mathrm{IA}}\in\{0,1\}$ denote the binary label.

The area under the precision--recall curve is
\[
\mathrm{AUPRC}=\int_0^1 \mathrm{Precision}(r)\,dr.
\]

Precision and recall at threshold $\tau$ are
\[
\mathrm{Precision}(\tau)=
\frac{\mathrm{TP}(\tau)}
{\mathrm{TP}(\tau)+\mathrm{FP}(\tau)},
\qquad
\mathrm{Recall}(\tau)=
\frac{\mathrm{TP}(\tau)}
{\mathrm{TP}(\tau)+\mathrm{FN}(\tau)}.
\]

AUROC is
\[
\mathrm{AUROC}=\Pr(p_i^{+}>p_j^{-}),
\]
where $p_i^{+}$ is the predicted risk for a failed initial attack event and $p_j^{-}$ is the predicted risk for a successful initial attack event.

For threshold-based reporting, the operating threshold is selected on the validation set:
\[
\tau^\star=
\arg\max_{\tau}
\frac{
2\,\mathrm{Precision}_{\mathrm{val}}(\tau)\,\mathrm{Recall}_{\mathrm{val}}(\tau)
}{
\mathrm{Precision}_{\mathrm{val}}(\tau)+\mathrm{Recall}_{\mathrm{val}}(\tau)
}.
\]
The predicted binary label is then
\[
\hat{y}_i^{\mathrm{IA}}
=
\mathbf{1}[p_i^{\mathrm{IA}}\geq \tau^\star].
\]

Balanced accuracy is
\[
\mathrm{BalancedAcc}
=
\frac{1}{2}
\left(
\frac{\mathrm{TP}}{\mathrm{TP}+\mathrm{FN}}
+
\frac{\mathrm{TN}}{\mathrm{TN}+\mathrm{FP}}
\right).
\]

For top-$k$ operational triage metrics, let $\mathcal{S}_k$ be the top $k\%$ of test events ranked by predicted failure probability. 
Then
\[
\mathrm{Recall@}k\%
=
\frac{
\sum_{i\in\mathcal{S}_k} y_i^{\mathrm{IA}}
}{
\sum_{i\in\mathcal{D}_{\mathrm{test}}^{\mathrm{IA}}} y_i^{\mathrm{IA}}
},
\qquad
\mathrm{Precision@}k\%
=
\frac{
\sum_{i\in\mathcal{S}_k} y_i^{\mathrm{IA}}
}{
|\mathcal{S}_k|
}.
\]

Brier score is
\[
\mathrm{Brier}
=
\frac{1}{n}
\sum_{i=1}^{n}
(p_i^{\mathrm{IA}}-y_i^{\mathrm{IA}})^2.
\]

Binary cross-entropy is
\[
\mathrm{BCE}
=
-\frac{1}{n}
\sum_{i=1}^{n}
\left[
y_i^{\mathrm{IA}}\log p_i^{\mathrm{IA}}
+
(1-y_i^{\mathrm{IA}})\log(1-p_i^{\mathrm{IA}})
\right].
\]

Expected calibration error is
\[
\mathrm{ECE}
=
\sum_{m=1}^{M}
\frac{|B_m|}{n}
\left|
\mathrm{acc}(B_m)-\mathrm{conf}(B_m)
\right|,
\]
where $B_m$ is the $m$-th confidence bin, $\mathrm{acc}(B_m)$ is the observed positive frequency in that bin, and $\mathrm{conf}(B_m)$ is the average predicted confidence.

\noindent\textbf{Task 2 regression metrics.}
For the containment-time task, let $h_i$ denote true containment hours, $\hat{h}_i$ denote predicted containment hours after converting from log space, and $\hat{y}_i^{\mathrm{TTC}}$ denote the predicted log-space target.

Mean absolute error in hours is
\[
\mathrm{MAE}_{h}
=
\frac{1}{n}
\sum_{i=1}^{n}
|\hat{h}_i-h_i|.
\]

Root mean squared error in hours is
\[
\mathrm{RMSE}_{h}
=
\sqrt{
\frac{1}{n}
\sum_{i=1}^{n}
(\hat{h}_i-h_i)^2
}.
\]

Median absolute error in hours is
\[
\mathrm{MedianAE}_{h}
=
\mathrm{median}_{i}
\left(
|\hat{h}_i-h_i|
\right).
\]

Log-space MAE is
\[
\mathrm{MAE}_{\log}
=
\frac{1}{n}
\sum_{i=1}^{n}
|\hat{y}_i^{\mathrm{TTC}}-y_i^{\mathrm{TTC}}|.
\]

Log-space RMSE is
\[
\mathrm{RMSE}_{\log}
=
\sqrt{
\frac{1}{n}
\sum_{i=1}^{n}
(\hat{y}_i^{\mathrm{TTC}}-y_i^{\mathrm{TTC}})^2
}.
\]

Spearman correlation is
\[
\rho_s
=
\mathrm{corr}
\left(
\mathrm{rank}(\hat{h}),
\mathrm{rank}(h)
\right).
\]

The coefficient of determination is
\[
R^2
=
1-
\frac{
\sum_{i=1}^{n}
(h_i-\hat{h}_i)^2
}{
\sum_{i=1}^{n}
(h_i-\bar{h})^2
}.
\]

\section{Related Work}\label{sec:related}

\noindent\textbf{Initial attack, suppression effectiveness, and suppression difficulty.}
Initial attack is usually evaluated as a suppression-system outcome, and prior studies show that the definition of ``success'' depends on containment objective, final-size threshold, response standard, and agency context~\cite{korkola2024ia,xu2023initialattack,butler2025victoria,katuwal2018predictattack}. 
Accordingly, operational studies model initial attack with variables that are not always available in public national data, including response time, resource deployment, station capacity, aircraft use, and local suppression objectives~\cite{lee2013deploying,katuwal2018predictattack,rashidi2018attacker,cardil2025suppression}. 
Such work also motivates our access and terrain features, because suppression difficulty is affected by fire behavior, terrain difficulty, roads, response opportunities, and suppression capacity rather than by weather alone~\cite{rodriguez2020suppression,cardil2025suppression,finney1998farsite,finney2006flammap}. 
\benchmarkname{} differs from these studies by fixing a public event-level contract across the contiguous United States and by excluding operational variables that are not consistently available near discovery time~\cite{short2017fpafod,pourmohamad2024fpafodattributes,korkola2024ia}.

\noindent\textbf{Containment duration and post-discovery outcomes.}
Containment time is related to initial attack failure, but it is not the same target: a fire can escape the first response and still be contained quickly, or it can remain active because of terrain, fuel, weather, resource limits, or reporting conventions~\cite{bhardwaj2025containment,rodriguez2020suppression,cardil2025suppression}. 
Machine-learning work on containment duration therefore treats the target as a continuous or transformed time variable rather than as a binary escape label~\cite{bhardwaj2025containment}. 
Due to that distinction, \benchmarkname{} reports containment time as an auxiliary regression task and keeps the initial attack failure task as the primary ranking problem~\cite{brier1950verification,guo2017calibration,korkola2024ia}.

\noindent\textbf{Public fire records and environmental covariates.}
FPA-FOD provides a national event backbone with discovery time, location, cause, final size, and containment fields, while recent FPA-FOD-Attributes work demonstrates the value of attaching physical, biological, social, and administrative covariates to fire occurrence records~\cite{short2017fpafod,pourmohamad2024fpafodattributes}. 
Weather and fire danger are commonly derived from gridded meteorological or reanalysis products such as gridMET, ERA5, and MERRA-2, while vegetation, fuel, and terrain features are commonly derived from LANDFIRE and related landscape data~\cite{abatzoglou2013gridmet,hersbach2020era5,gelaro2017merra2,landfire,finney1998farsite,finney2006flammap}. 
Similarly, access and exposure variables can be attached from vector and demographic products such as OpenStreetMap and WorldPop, but these sources have different spatial resolution, update frequency, and missingness patterns~\cite{osm,worldpop,chicas2022systematic,shmuel2022global}. 
Thus, our canonicalization pipeline follows the public-data integration direction of prior occurrence and susceptibility work while changing the prediction target from ignition or susceptibility to initial attack outcome~\cite{singla2021wildfiredb,shen2023wildfirefpa,shmuel2022global,pourmohamad2024fpafodattributes}.

\noindent\textbf{Remote sensing for active fire detection, burned area, and fire mapping.}
Satellite products are central to modern wildfire monitoring, but they measure thermal activity or burned-area evidence rather than initial attack success~\cite{schroeder2014viirs,giglio2016modis,nasa_firms,mtbs}. 
Deep-learning studies and reviews have covered active-fire detection, smoke detection, burned-area segmentation, and satellite-based fire mapping, and these tasks often use image-level or pixel-level labels rather than incident-level suppression outcomes~\cite{ghali2023deepreview,ghali2023rsreview,smokeviz2025,tssatfire2025benchmark}. 
Because small or newly discovered fires may lack a matched VIIRS detection because of orbital timing, cloud, smoke, geolocation uncertainty, or intensity limits, \benchmarkname{} treats discovery-day VIIRS as an optional feature and retains events without matched detections~\cite{schroeder2014viirs,giglio2016modis,korkola2024ia}. 
This choice prevents the benchmark from becoming a satellite-detection benchmark and keeps the sample unit aligned with the fire report~\cite{short2017fpafod,nasa_firms,pourmohamad2024fpafodattributes}.

\noindent\textbf{Wildfire spread prediction and fire simulators.}
Physics-based and coupled fire models such as FARSITE, FlamMap, ForeFire, and WRF-Fire represent fire growth, spread, or fire-atmosphere interaction under explicit environmental assumptions~\cite{finney1998farsite,finney2006flammap,filippi2014forefire,coen2011wrfsfire}. 
Data-driven spread benchmarks such as WildfireDB, Next Day Wildfire Spread, WildfireSpreadTS, WSTS+, and Sim2Real-Fire instead evaluate models that predict future fire masks, burned cells, or spread trajectories from remote-sensing and environmental grids~\cite{singla2021wildfiredb,huot2022nextday,gerard2023wildfirespreadts,lahrichi2025wstsplus,sim2realfire2024}. 
These resources are important for fire-growth modeling, yet their target is future spatial propagation rather than whether an incident remains below an initial attack size objective~\cite{radke2019firecast,firepred2023hybrid,asufm2024vit,cnnaspp2024xai}. 
Accordingly, \benchmarkname{} complements spread benchmarks by using an event-level label, a chronological split, and a forbidden-feature policy built around discovery-time suppression prediction~\cite{korkola2024ia,xu2023initialattack,butler2025victoria}.

\noindent\textbf{Models for tabular, temporal, spatial, and spatiotemporal inputs.}
Initial attack prediction requires comparing models across heterogeneous input forms, because event reports, weather sequences, landscape patches, and patch sequences encode different parts of the same incident~\cite{short2017fpafod,abatzoglou2013gridmet,landfire,osm,worldpop}. 
For tabular data, logistic regression, support-vector machines, random forests, gradient boosting, LightGBM, XGBoost, and multilayer perceptrons are common structured-data baselines, and probabilistic evaluation further requires calibration-aware metrics~\cite{cox1958binary,cortes1995svm,breiman2001random,ke2017lightgbm,chen2016xgboost,rumelhart1986backprop,guo2017calibration,brier1950verification,lakshminarayanan2017simple}. 
For temporal data, recurrent, convolutional, and attention-based sequence models provide complementary ways to encode short weather histories, while weather and Earth-system forecasting benchmarks motivate careful handling of spatiotemporal covariates~\cite{hochreiter1997lstm,bai2018tcn,vaswani2017attention,gao2022earthformer,prithviwxc2024,chaosbench2024,seasonbenchea2025}. 
For spatial and spatiotemporal patches, convolutional encoders, U-Net-style decoders, attention U-Nets, DeepLab-style decoders, transformer segmentation backbones, ConvLSTM/ConvGRU units, PredRNN, TrajGRU, MAU, UTAE, Rainformer, and SwinLSTM-style modules provide established templates for processing gridded or video-like environmental inputs~\cite{he2016deep,ronneberger2015unet,oktay2018attention,chen2018deeplab,cao2021swinunet,xie2021segformer,shi2015convlstm,ballas2016convgru,wang2022predrnn,shi2017trajgru,chang2021mau,garnot2021utae,bai2022rainformer,tang2023swinlstm}. 
Our baseline suite adapts these architectures to scalar event-level prediction by replacing dense prediction heads with global pooling and a scalar output, thereby keeping the comparison focused on the same initial attack and containment labels~\cite{long2015fcn,strudel2021segmenter,zheng2021setr,dosovitskiy2021vit,wu2024earthfarsser}.

\begin{table*}[t]
\centering
\small
\caption{Representative prior work and how \benchmarkname{} differs. The table groups related work by target, because the target determines the valid sample unit, split, and leakage constraints.}
\label{tab:ia_related_work}
\resizebox{\textwidth}{!}{%
\begin{tabular}{llll}
\toprule
\textbf{Line of work} & \textbf{Examples} & \textbf{Typical target} & \textbf{Difference from \benchmarkname} \\
\midrule
Initial attack outcome & Korkola et al.; Xu et al.; Butler et al.~\cite{korkola2024ia,xu2023initialattack,butler2025victoria} & IA success or escape & Regional or agency-specific definitions; no shared U.S. multi-modal cache \\
Suppression difficulty & Lee et al.; Katuwal et al.; Cardil et al.~\cite{lee2013deploying,katuwal2018predictattack,cardil2025suppression} & Response success or difficulty & Often uses operational variables not consistently public nationally \\
Containment duration & Bhardwaj~\cite{bhardwaj2025containment} & Time to containment & Regression target; not a primary IA failure benchmark \\
Fire occurrence and susceptibility & FPA-FOD-Attributes; WildfireDB; global susceptibility maps~\cite{pourmohamad2024fpafodattributes,singla2021wildfiredb,shmuel2022global} & Ignition, occurrence, or susceptibility & Predicts where fires occur, not whether a discovered fire escapes IA \\
Active fire and burned-area mapping & VIIRS/MODIS/MTBS and remote-sensing DL reviews~\cite{schroeder2014viirs,giglio2016modis,mtbs,ghali2023deepreview,ghali2023rsreview} & Thermal detection or burned pixels & Sensor or pixel labels, not incident-level suppression outcomes \\
Wildfire spread benchmarks & FireCast, Next Day Wildfire Spread, WildfireSpreadTS, WSTS+, Sim2Real-Fire~\cite{radke2019firecast,huot2022nextday,gerard2023wildfirespreadts,lahrichi2025wstsplus,sim2realfire2024} & Future fire mask or spread trajectory & Forecasts spatial propagation rather than initial attack failure \\
\bottomrule
\end{tabular}%
}
\end{table*}

\section{Discussion and Limitations}
\label{sec:discuss}

\paragraph{FPA-FOD events, not FIRMS detections, define the benchmark.}
A central design choice in \benchmarkname{} is that the sample unit is the FPA-FOD natural wildfire event, not the FIRMS/VIIRS active-fire detection. 
This choice is necessary because the benchmark target is an event-level initial attack outcome: whether a reported wildfire ultimately remains small or escapes initial attack. 
FIRMS/VIIRS provides valuable discovery-day thermal evidence, but it does not provide a complete event catalog, a final fire size, or a containment outcome. 
Therefore, we treat FIRMS/VIIRS as an input source attached to FPA-FOD events rather than as the dataset definition.

\begin{figure*}[t]
\centering
\includegraphics[width=1\textwidth]{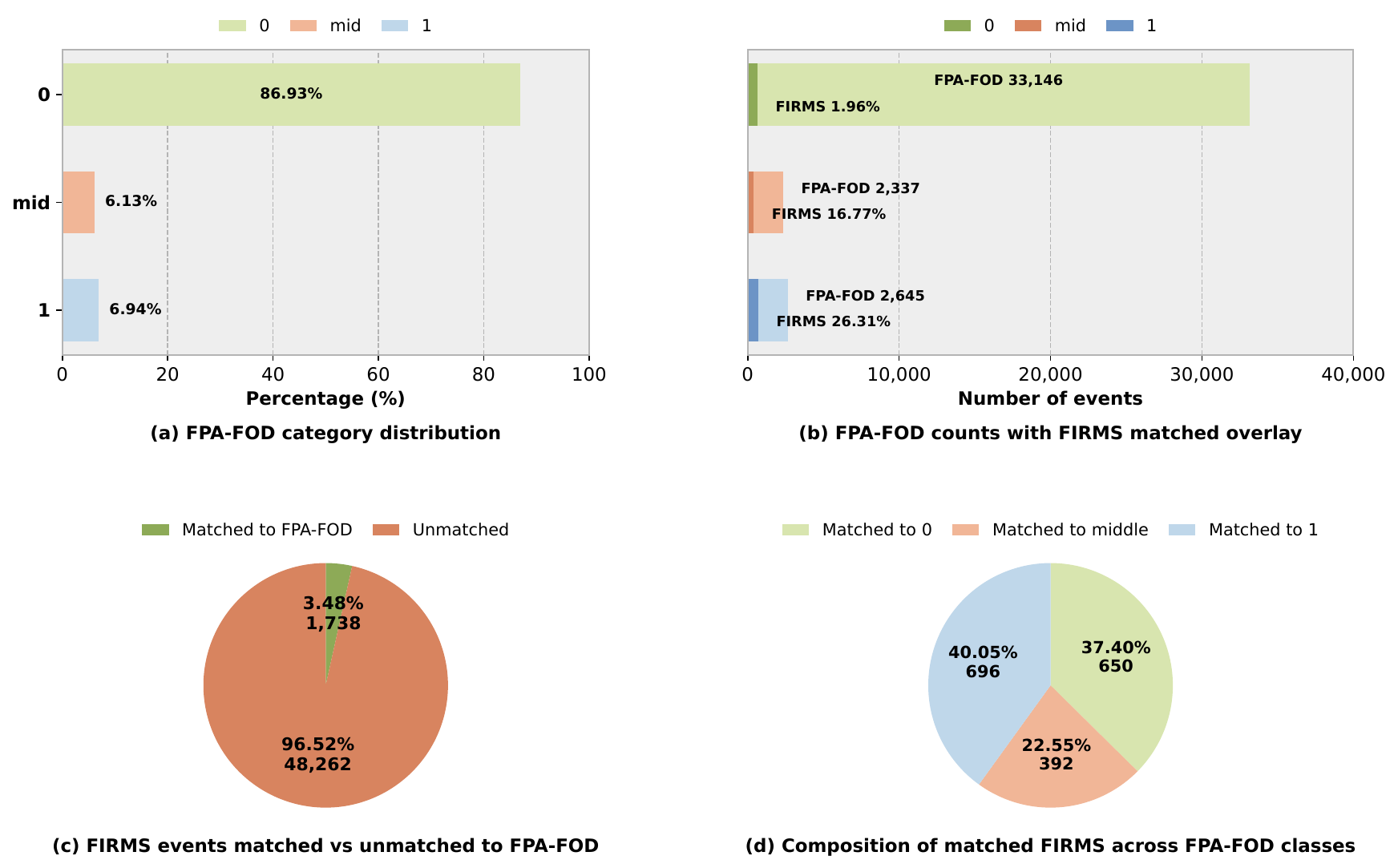}
\caption{
FIRMS/VIIRS matches are a small and biased subset of FPA-FOD natural wildfire events. 
The full FPA-FOD event set is highly imbalanced, whereas FIRMS-matched events are closer to balanced between success and failure. 
Thus, FIRMS/VIIRS is used as an input feature source, not as the benchmark sample definition.
}
\label{fig:figure2}
\end{figure*}

Figure~\ref{fig:figure2} illustrates why this distinction matters. 
The FPA-FOD event distribution is highly imbalanced: the small-fire / initial attack success class accounts for most retained natural wildfire events, while escaped fires are rare. 
However, discovery-day FIRMS/VIIRS matching is much more frequent for larger or more extreme events. 
Only a small fraction of class-0 events have matched discovery-day FIRMS/VIIRS evidence, whereas the matched fraction is substantially higher for intermediate and escaped-fire events. 
Thus, restricting the benchmark to FIRMS-matched events changes the event population and makes the label distribution much less rare-event-like. 
A model trained and evaluated only on this subset can appear much more accurate, but it is solving an easier and less representative problem.

\paragraph{FIRMS-detected subset diagnostic.}
We additionally define a diagnostic subset to understand the effect of conditioning the benchmark on discovery-day satellite observability. 
Let
\[
z_i^{R}
=
\mathbbm{1}\{\mathrm{has\_viirs\_detection}_{i,D}=1\},
\qquad
\mathcal{I}_{R}
=
\{i:z_i^{R}=1\}.
\]
On this subset, we evaluate a FIRMS-focused contract using only FPA-FOD metadata and discovery-day VIIRS features:
\[
(M_i,R_i)
\rightarrow
y_i^{\mathrm{IA}},
\qquad
i\in\mathcal{I}_{R}.
\]
This diagnostic does not redefine the label: the target remains the FPA-FOD-derived initial attack-failure label \(y_i^{\mathrm{IA}}\). 
The only change is that the evaluated event set is restricted to FPA-FOD events with matched discovery-day FIRMS/VIIRS evidence.

\begin{table}[t]
\centering
\small
\caption{
Prediction-count diagnostic on the FIRMS-detected test subset. 
The subset contains only FPA-FOD events with matched discovery-day VIIRS detections. 
Predicted counts are averaged over five seeds.
}
\label{tab:firms_detected_prediction_counts}
\begin{tabular}{lccccc}
\toprule
\textbf{Model} & \textbf{Test N} & \textbf{True 0} & \textbf{True 1} & \textbf{Pred 0} & \textbf{Pred 1} \\
\midrule
XGBoost     & 303 & 142 & 161 & 78.0\pmstd{7.8}   & 225.0\pmstd{7.8} \\
Transformer & 303 & 142 & 161 & 0.2\pmstd{0.4}    & 302.8\pmstd{0.4} \\
Swin-UNet   & 303 & 142 & 161 & 144.6\pmstd{13.7} & 158.4\pmstd{13.7} \\
SwinLSTM    & 303 & 142 & 161 & 105.4\pmstd{63.8} & 197.6\pmstd{63.8} \\
\bottomrule
\end{tabular}
\end{table}

Table~\ref{tab:firms_detected_prediction_counts} shows why this subset is diagnostic rather than primary. 
The FIRMS-detected test subset contains 303 events, with escaped fires forming the majority class: 161 of 303 events, or 53.1\%. 
This class balance is very different from the full benchmark, where initial attack failure is a rare event. 
Therefore, FIRMS-only results cannot be directly compared with the full benchmark leaderboard. 
They answer a conditional question: once a reported wildfire is already visible to VIIRS on the discovery day, how much can the matched thermal signal separate successful initial attack from escaped fires?

This distinction is important for benchmark design. 
Using FIRMS/VIIRS detection as the event definition would bias the benchmark toward fires that are hotter, larger, more persistent, or easier to observe from orbit. 
It would exclude many reported FPA-FOD events without matched discovery-day detections, including small or quickly contained fires and events missed because of satellite timing, cloud, smoke, or detection limits. 
For this reason, \benchmarkname{} keeps FPA-FOD as the event backbone and uses FIRMS/VIIRS only as an optional discovery-day input feature. 
The FIRMS-detected subset is useful for analyzing satellite-observable fires, but it should not replace the full event-level benchmark.

\paragraph{Initial attack failure remains a rare-event prediction problem.}
The full FPA-FOD-defined benchmark preserves the operational difficulty of initial attack prediction: escaped fires are rare, and many reported fires remain small. 
This is why we use AUPRC as the primary metric and report calibration metrics in addition to thresholded classification scores. 
High accuracy on the full benchmark would be easy to obtain by predicting the majority class, but such a model would be useless for identifying fires that require escalation. 
The benchmark therefore emphasizes rare-failure prioritization rather than aggregate correctness.

\paragraph{Containment duration is a harder auxiliary target.}
The auxiliary containment-duration analysis shows that the same discovery-time public source stack contains signal about time to containment, but only partially explains the target. 
The best models achieve modest \(R^2\), around 0.18, indicating that early public variables explain a limited fraction of containment-duration variance. 
This result is consistent with the nature of the target. 
Containment duration depends not only on early weather, fuel, topography, access, population context, and satellite fire signal, but also on suppression tactics, crew and aircraft availability, changing weather after discovery, incident command decisions, reporting conventions, and containment criteria. 
Many of these drivers are not fully observed in our public discovery-time input stack. 
We therefore interpret containment-duration prediction as an auxiliary stress test rather than as the primary benchmark objective.

\paragraph{Limitations of public event records.}
FPA-FOD provides a nationwide event backbone, but it is still a reporting-derived dataset. 
Reported discovery locations may be approximate, final fire size may be recorded with agency-specific conventions, and containment timestamps may be missing or inconsistently reported. 
Our thresholded initial attack label reduces some ambiguity by separating clearly small fires from clearly escaped fires, but it does not eliminate label noise. 
Events in the intermediate 10--50 ha range are marked missing for the binary initial attack target to avoid forcing ambiguous cases into either class.

\paragraph{Limitations of source alignment.}
All non-FPA-FOD sources are aligned to events through reported discovery date, location, and year. 
This makes the benchmark reproducible, but it also introduces uncertainty. 
FIRMS/VIIRS detections depend on overpass timing, cloud and smoke conditions, sensor geometry, and the strictness of the matching radius. 
gridMET represents coarse weather fields rather than exact fireground conditions. 
LANDFIRE fuel and vegetation layers are treated as static landscape context, which does not fully capture annual disturbance, treatment, or fuel-moisture dynamics. 
OSM roads and fire stations are public proxies for access and response proximity, not direct measurements of dispatch decisions, travel time, crew availability, or suppression capacity. 
WorldPop provides human-settlement context but does not directly encode evacuation priority, asset value, or wildland-urban-interface structure.

\paragraph{Scope of the current benchmark.}
The current version focuses on natural wildfire events from 2016--2020 and uses a fixed chronological split. 
This scope reduces ignition-source heterogeneity and supports future-year evaluation, but it does not cover all human-caused wildfire dynamics. 
Extending the benchmark to all-cause wildfires, adding explicit suppression-resource records, incorporating post-discovery weather evolution, or using high-resolution perimeter growth would enable richer operational questions. 
However, those additions would also change the information contract. 
The present benchmark intentionally focuses on what can be attached to a reported natural wildfire near discovery time using reproducible public data.

\section{Reproducibility Artifacts}
The repository is organized around three scripts. \texttt{pipeline.py} builds canonical data from raw sources, \texttt{dataloader.py} builds tabular, temporal, spatial, and spatiotemporal caches from canonical outputs, and \texttt{train.py} trains the baseline models. Experiment directories store the run configuration, model history, metrics, validation and test predictions, and checkpoints for neural models. Summary scripts aggregate metrics by task, representation, model, and seed. This trace is meant to make the benchmark easy to inspect: every reported number should be recoverable from saved \texttt{metrics.json} files and prediction parquet files.

\section{LLM Usage}
\label{app:llm_usage}

Large language models were used only for writing assistance, including grammar checking, wording refinement, formatting suggestions, and readability editing. They were not used to generate the core methodology, design experiments, produce experimental results, create evaluation labels, or make routing decisions. All technical claims, mathematical formulations, experimental settings, and reported results were checked and finalized by the authors.

\end{document}